\documentclass[usenatbib]{mn2e}
\usepackage{graphicx,epsfig}
\usepackage{subfigure}

\def\apj{{\em ApJ}}
\def\apjs{{\em ApJS}}
\def\apjl{{\em ApJ}}
\def\aap{{\em A\&A}}
\def\aj{{\em AJ}}

\def\mnras{{\em MNRAS}}

\def\araa{Annual Review of Astron and Astrophys}
 
\def\aaps{{\em A\&AS}} 

\newcommand{\lunits}{\ensuremath{\,\rmn{erg\,s^{-1}\,Hz^{-1}}}}  
\newcommand{\rlunits}{\ensuremath{\,\rmn{erg\,s^{-1}}}}          
\newcommand{\h}{\ensuremath{\rmn{^h}}}                           
\newcommand{\m}{\ensuremath{\rmn{^m}}}                           
\newcommand{\s}{\ensuremath{\rmn{^s}}}                           
\newcommand{\mum}{\,\mbox{\ensuremath{\rmn{\mu m}}}}             
\newcommand{\mjy}{\,\mbox{\ensuremath{\rmn{mJy}}}}               
\newcommand{\mujy}{\,\mbox{\ensuremath{\rmn{\mu Jy}}}}           
\newcommand{\jyb}{\,\mbox{\ensuremath{\rmn{Jy/beam}}}}           
\newcommand{\mjyb}{\,\mbox{\ensuremath{\rmn{mJy/beam}}}}         
\newcommand{\mujyb}{\,\mbox{\ensuremath{\rmn{\mu Jy/beam}}}}     
\newcommand{\uv}{\mbox{{\it uv}-}}                                                  
\newcommand{\hii}{\mbox{H{\footnotesize{II}}}}                                           
\newcommand{\snrate}{\ensuremath{\nu_{\rmn{\mbox{\tiny SN}}}}}                           
\newcommand{\speak}{\ensuremath{S_{\rmn{peak}}}}                                         
\newcommand{\stot}{\ensuremath{S_{\rmn{tot}}}}                                           
\newcommand{\lir}{\ensuremath{L_{\rmn{IR}}}}                                             
\newcommand{\lfir}{\ensuremath{L_{\rmn{FIR}}}}                                           
\newcommand{\alphtot}{\ensuremath{\alpha_{\rmn{tot}}^{\rmn{\mbox{\tiny L-C}}}}} 
\newcommand{\alphpeak}{\ensuremath{\alpha_{\rmn{peak}}^{\rmn{\mbox{\tiny L-C}}}}} 
\newcommand{\alphpp}{\ensuremath{\alpha_{\rmn{\mbox{\tiny p-p}}}^{\rmn{\mbox{\tiny L-C}}}}} 
\newcommand{\tb}{\ensuremath{T_{\rmn{B}}}}                                               

\title[EVN observations towards IRAS\,23365$+$3604]
{EVN observations of the farthest and brightest ULIRGs in the local Universe: the case of IRAS\,23365$+$3604}
\author[C.~Romero-Ca\~nizales et al.]
  {Cristina Romero-Ca\~nizales,$^{1,2}$\thanks{E-mail: crroca@utu.fi}
  Miguel \'Angel P\'erez-Torres,$^1$
  \& Antxon Alberdi$^1$ \\ 
$^1$Instituto de Astrof\'{\i}sica de Andaluc\'{\i}a -- CSIC, PO Box 3004, 18080 Granada, Spain \\
$^2$Tuorla Observatory, Department of Physics and Astronomy, University of Turku, V\"ais\"al\"antie 20, FI-21500 
    Piikki\"o, Finland}

\begin{document}

\date{Accepted 2012 January 24. Received 2012 January 24; in original form 2011 November 4}
\pagerange{\pageref{firstpage}--\pageref{lastpage}} \pubyear{2011}

\maketitle

\label{firstpage} 

\begin{abstract} 
We present high-resolution, high-sensitivity radio images of the ultra-luminous infrared galaxy (ULIRG) IRAS\,23365$+$3604. 
We performed contemporaneous observations at 1.7 and 5.0\,GHz, in three epochs separated by one year from each other, with 
the European very long baseline interferometry Network (EVN). We also present complementary Multi-Element Radio Linked 
Interferometry Network (MERLIN) at 1.6 and 5.0\,GHz, and archival Very Large Array (VLA) data, taken at 1.4 and 4.9\,GHz. 
We find that the emission at $\sim$5.0\,GHz remains quite compact as seen at different resolutions, whereas at $\sim$1.7\,GHz, 
high resolution imaging reveals some extended structure. The nuclear region has an approximate linear size of 200\,pc and shows
the presence of two main emission components: i) one with a composite spectrum due to ongoing non-thermal activity (probably 
due to recently exploded supernovae and AGN activity), ii) another one with a steep spectrum, likely dominated by an old population 
of radio emitters, such as supernova remnants (SNRs). Radiative losses are important, so re-acceleration or replenishment of new 
electrons is necessary. We estimate a  magnetic field strength of $\sim18\,\mu$G at galactic, and $\sim175\,\mu$G at nuclear scales, 
which are typical for galaxies in advanced mergers. 

\end{abstract}


\section{Introduction}\label{sec:introulirg}

Galaxies with very high infrared (IR) luminosities ($\lir=L[8$--$1000\,\mum]>10^{12}$\,L$_{\sun}$)
known as ultra luminous IR galaxies (ULIRGs), dominate the IR background and the star
formation rate (SFR) density at z$\sim$2 \citep{ulirgz}. Although uncommon at lower redshifts, the presence
of ULIRGs in the local Universe offers the opportunity of investigating their parsec scale structure,
while profiting from the high angular resolution provided by current instrumentation. Studying ULIRGs in the 
local Universe is of great importance since it can aid to understand their high-redshift analogues which 
dominate the sub-mm sky \citep[see e.g.,][]{lilly99}.

It is thought that ULIRGs represent a key stage in the formation of optical quasi-stellar objects (QSOs) and 
powerful radio galaxies \citep[e.g.,][]{sanders88}. A study based on {\it HST} observations and {\it N}-body
simulations point to diverse evolutionary paths, not necessarily including a QSO phase \citep{farrah01}. There is 
however a general agreement on gas-rich galaxy merging as the origin of ULIRGs \citep{sanders}, and on the ubiquity 
of enhanced star-formation, which can be found in combination with different flavours of active galactic nucleus 
(AGN) activity \citep[e.g.,][]{farrah03}. Which of these two contributions dominates and is primarily responsible 
for the overall dust heating, is still an open question.  

\begin{table*}
\centering
\begin{minipage}{168mm}
\caption{\protect{The EVN ULIRG sample. Values in columns 2--6 are those provided by \citet{sanders03}. The supernova 
          rates in column 7 were obtained following the empirical relation between CCSN rate and $L[8$--$1000\mum]$ 
          for starburst galaxies obtained by \citet{seppo01}, without discounting AGN contribution, if any. In column 
          8 we show the flux density range a SN would display when having a typical type II and type IIn SN peak 
          luminosity, denoted by $L \sim 10^{27-28}$\,\lunits{}.}}\label{tab:evnsample}
\begin{tabular}{lllccccc} \hline
 \multicolumn{1}{c}{IRAS name} & \multicolumn{2}{c}{{\it IRAS} position (J2000)}&
   \multicolumn{1}{c}{Distance} & \multicolumn{1}{c}{Redshift} & \multicolumn{1}{c}{log$_{10}(\lir{}/\rmn{L}_{\sun}$)} 
   &  \multicolumn{1}{c}{\snrate{}}  &  \multicolumn{1}{c}{$S_{\nu}^{\rmn{SN}}$} \\
    \multicolumn{1}{c}{ } & \multicolumn{1}{c}{$\alpha$(\h{}\,\m{}\,\s{})} & \multicolumn{1}{c}{$\delta$(\degr{}\,\arcmin{}\,\arcsec{})} 
& \multicolumn{1}{c}{(Mpc)} & \multicolumn{1}{c}{ } & \multicolumn{1}{c}{ } & \multicolumn{1}{c}{(yr$^{-1}$)} & \multicolumn{1}{c}{(\mujy{})} \\
\multicolumn{1}{c}{(1)} & \multicolumn{1}{c}{(2)} & \multicolumn{1}{c}{(3)} & \multicolumn{1}{c}{(4)} & \multicolumn{1}{c}{(5)} 
  & \multicolumn{1}{c}{(6)}  & \multicolumn{1}{c}{(7)}  & \multicolumn{1}{c}{(8)} \\
\hline
07251$-$0248   &  07 27 37.5  &  $-$02 54 55  & 344 & 0.088  & 12.32  &  5.6  &  7--71 \\
19297$-$0406   &  19 32 22.1  &  $-$04 00 02  & 338 & 0.086  & 12.37  &  6.3  &  7--73 \\
19542$+$1110   &  19 56 35.4  &  $+$11 19 03  & 257 & 0.065  & 12.04  &  3.0  &  13--127 \\
23365$+$3604   &  23 39 01.7  &  $+$36 21 14  & 252 & 0.064  & 12.13  &  3.6  &  13--132 \\
    \hline
   \end{tabular} 
 \end{minipage}
\end{table*}

\citet{kewley06} presented a robust classification scheme of galaxies based on optical emission line ratios of a large sample
of galaxies from the Sloan Digital Sky Survey (SDSS). This scheme allows to discriminate between starbursts, Seyferts, 
low-ionization narrow emission-line regions (LINERs), and composite starbursts-AGN types. More recently, \citet{yuan10} 
used the \citeauthor{kewley06} scheme to classify a sample of IR selected galaxies, as a function of IR luminosity and merger
stage. Their results support an evolutionary scenario in which ULIRGs are dominated by starburst activity at an early 
merger stage; at intermediate stages, ULIRGs would be powered by a composite of starburst-AGN activity; and finally, at 
later stages, an AGN would dominate the emission. 

A key feature of ULIRGs is their large dust content, which is heated by a central power source, or sources. Since optical 
obscuration is high, radio observations (i.e., extinction free) represent the most direct way to distinguish between a 
starburst and an AGN, via the detection of supernovae (SNe), supernova remnants (SNRs) and/or compact sources at mas-resolution
with a high brightness temperature (\tb{}), possibly accompanied by a core-jet morphology and usually associated with a high X-ray
luminosity. Very long baseline interferometry (VLBI) observations have been particularly useful, for instance, to 
discover a population of bright radio SNe and SNRs in the nuclear regions of the ULIRGs Arp\,220 \citep{smith98} and Mrk\,273 
\citep{carilli00}. This has also been the case for LIRGs ($L_{\rmn{IR}}>10^{11}$\,L$_{\sun}$), such as Arp\,299 where a prolific 
starburst and a low-luminosity AGN (LLAGN) were discovered \citep[][respectively]{neff, paper2} through VLBI observations, or the 
recent detection of AGN activity in a number of LIRGs from the Compact Objects in Low-power AGN (COLA) sample \citep{parra10}.

\section{The EVN ULIRG sample}
\label{sec:samp}

This is the first of a series of papers presenting European VLBI Network (EVN) high-resolution, high-sensitivity images 
of a sample of four of the farthest and brightest ultra luminous infrared galaxies (ULIRGs) in the local Universe 
($\rmn{z}<0.1$), part of the project entitled ``The dominant heating mechanism in the central regions of
ULIRGs'' (PI: P\'erez-Torres).

The sample of ULIRGs we present results from the following selection 
process. We have first selected those sources from the {\it IRAS} Revised Bright Galaxy Sample \citep{sanders03} 
having log$_{10}(\lir{}/\rmn{L}_{\sun})>12$, from which large supernova rates (\snrate{}) were expected. 
We further constrained our sample by selecting those objects with $\delta>-5\degr{}$ 
(in order to obtain a good \uv{}coverage with the EVN), which also appear in the 1.4\,GHz Atlas Catalogue of the 
{\it IRAS} Bright Galaxy Sample \citep{condon90,condon96}, as to ensure their radio emission detection. Finally, we
selected those ULIRGs for which neither Multi-Element Radio Linked Interferometry Network (MERLIN) nor deep VLBI data 
existed in the literature, and for which MERLIN or EVN archival data are not available. The resulting sample contains four 
of the brightest and farthest ULIRGs in the local Universe (Table \ref{tab:evnsample}), for which we aimed to unveil their 
dominant heating mechanisms.

The needed rms to obtain 3$\sigma$ detections of typical type II core-collapse SNe (CCSNe) in the most distant ULIRGs of 
our sample, is quite low ($\sim 2\mujy{}$, for peak luminosities $\sim 10^{27}$\,\lunits) and therefore are well below our 
detection limit. Nevertheless, it is also expected that more luminous systems provide denser environments, which in turn 
favour the production of very luminous CCSNe \citep[e.g., type IIn SNe in Arp\,220;][]{parra}. Moreover, the more luminous 
a radio SN (RSN) is, the longer it will take for it to reach its peak brightness \citep[see figure 5 in][]{alberdi06}. For 
instance, the RSN A0 discovered in the nuclear region of Arp\,299 in 2003 \citep[see][]{neff}, is still detected after 
several years and remains particularly strong \citep{paper1,paper2}. A similar scenario in the dense nuclear 
regions of our sample of ULIRGs can be expected. Furthermore, in the circumnuclear regions of LIRGs and ULIRGs, we also 
expect SN activity to occur. Two remarkable examples are SN\,2000ft and SN\,2004ip, discovered at 600\,pc and 500\,pc from 
the nucleus of galaxies NGC\,7469 \citep{colina01, alberdi06, sn2000ftradio} and IRAS\,18293$-$3413 
\citep{sn2004ip, sn04ipradio}, respectively.

It is worth noting that the empirical relation between CCSN rate and $L[8$--$1000\mum]$ for starburst galaxies 
obtained by \citet{seppo01}:
\begin{equation}\label{eq:lum_nusnr}
\left(\frac{\snrate{}}{\rmn{yr}^{-1}}\right)=2.7\times10^{-12}\times\left(\frac{\lir{}}{\rmn{L}_{\sun}}\right),
\end{equation}
assumes no AGN contribution to the IR luminosity. The same is true with a similar relation that results from the combination 
of equations 20 and 26 in \citet{condon}, and which yields slightly larger values, i.e., 
\[\left(\frac{\snrate{}}{\rmn{yr}^{-1}}\right)\sim3.7\times10^{-12}\times\left(\frac{\lfir}{\rmn{L}_{\sun}}\right),\]
with $\lfir{}=L[40$--$400\,\mum]$\,L$_{\sun}$. Thus, if an AGN is present, the values for \snrate{} in Table \ref{tab:evnsample} 
represent upper limits, and a quantitative estimate of the AGN contribution to the IR luminosity is needed before deriving
reliable CCSN rates.

\subsection{The case of IRAS\,23365$+$3604}\label{sec:iras2336}

IRAS\,23365$+$3604 (hereafter IRAS\,23365) is thought to be in an advanced merger state \citep{sopp90}. There is no 
companion galaxy so far detected with either Very Large Array (VLA) or Two Micron All Sky survey (2MASS) observations 
\citep[see e.g.][]{sopp90, yuan10}. \citet{klaas91} report a companion candidate, a small galaxy located $\approx50$\,kpc 
(projected distance) away from IRAS\,23365, which is however not considered to be the cause of the extremely irregular and 
disturbed morphology of IRAS\,23365. The optical spectrum of this ULIRG seems to be the result of the superposition of LINER 
and \hii{}-region like components. Such AGN-starburst composite spectrum have been confirmed in other studies 
\citep[e.g.,][]{veron97, yuan10}. Chandra X-ray observations have evidenced the presence of an AGN (possibly Compton-thick) 
in this source \citep{iwasawa11}.

At a distance of 252\,Mpc (1~mas $\approx$ 1.2\,pc), the high luminosity of IRAS\,23365 (log$_{10}(\lir{}/$L$_{\sun})=12.13$) 
corresponds to a CCSN rate of $\approx3.6$\,yr$^{-1}$, according to Equation \ref{eq:lum_nusnr}. As indicated in section 
\ref{sec:samp}, this estimate does not consider an AGN contribution. According to \citet{farrah03}, the AGN contribution 
in IRAS\,23365 is approximately 35 per cent of the total \lir{}, and the rest is due to a starburst, from which we infer that
$\snrate{}\approx 2.4$\,yr$^{-1}$.

\begin{table*}
\centering
\begin{minipage}{168mm}
\caption{\protect{Parameters of the EVN observations. The stations (location, diameter) used in the different
observing runs are: Ef-Effelsberg (Germany, 100\,m), Wb-Westerbork array (NL, 14$\times$25\,m), Jb1-Lovell (UK, 76\,m), 
Jb2-MK\,II (UK, 25\,m), On-Onsala (Sweden, 25\,m), Mc-Medicina (Italy, 32\,m), Nt-Noto (Italy, 32\,m), Tr-Torun (Poland, 
32\,m), Ur-Urumqi  (China, 25\,m), Cm-Cambridge (UK, 32\,m), Kn-Knockin (UK, 25\,m), Ar-Arecibo (Puerto Rico, 305\,m), 
Ys-Yebes (Spain, 40\,m). In column 6 we show the phase reference source used in each epoch and in column 7 their associated
peak intensities.}} \label{tab:evnobs}
\begin{tabular}{cllclcc} \hline
Label & \multicolumn{1}{c}{Project} & \multicolumn{1}{c}{Observing}& \multicolumn{1}{c}{Frequency}
   &\multicolumn{1}{c}{Participating} & \multicolumn{1}{c}{Phase} & \multicolumn{1}{c}{\speak{}}  \\
 &     & \multicolumn{1}{c}{date}     & \multicolumn{1}{c}{(GHz)}    &\multicolumn{1}{c}{stations}     
    & \multicolumn{1}{c}{calibrator}   & \multicolumn{1}{c}{(\jyb{})} \\
\multicolumn{1}{c}{(1)} & \multicolumn{1}{c}{(2)} & \multicolumn{1}{c}{(3)} & \multicolumn{1}{c}{(4)} & \multicolumn{1}{c}{(5)} 
  & \multicolumn{1}{c}{(6)}  & \multicolumn{1}{c}{(7)}  \\
\hline
L1 & EP061A & 2008-02-29 & 1.7 & Ef, Wb, Jb1, On, Mc, Nt, Tr, Ur, Cm         &  J2333$+$3901  & 0.34 $\pm$ 0.02 \\
C1 & EP061C & 2008-03-11 & 5.0 & Ef, Wb, Jb1, On, Mc, Nt, Tr, Ur, Cm         &  J2333$+$3901  & 0.21 $\pm$ 0.01 \\
L2 & EP064D & 2009-03-07 & 1.7 & Ef, Wb, Jb2, On, Mc, Nt, Tr, Ur, Cm, Kn, Ar &  J2333$+$3901  & 0.48 $\pm$ 0.02 \\
C2 & EP064B & 2009-02-28 & 5.0 & Ef, Wb, Jb2, On, Mc, Nt, Tr, Ur, Cm, Kn, Ar &  J2333$+$3901  & 0.23 $\pm$ 0.01 \\
L3 & EP064J & 2010-03-08 & 1.7 & Ef, Wb, Jb1, On, Mc, Nt, Tr, Ur, Cm, Kn     &  J2330$+$3348  & 0.65 $\pm$ 0.03 \\
C3 & EP064L & 2010-03-20 & 5.0 & Ef, Wb, Jb1, On, Mc, Nt, Tr, Ur, Cm, Kn, Ys &  J2330$+$3348  & 0.64 $\pm$ 0.03 \\
    \hline
   \end{tabular} 
 \end{minipage}
\end{table*}

\section{EVN Observations and data reduction}
\label{sec:obsevn}

We performed EVN observations of IRAS\,23365 quasi-simultaneously at L- ($\nu \sim 1.7$\,GHz or 
$\lambda \sim 18$\,cm) and C-band ($\nu \sim 5$\,GHz or $\lambda \sim 6$\,cm) in three epochs with a time span 
among them of approximately one year (see Table \ref{tab:evnobs}).

\begin{table*}
\centering
\begin{minipage}{168mm}
\caption{\protect{EVN observational data, as measured from the images shown in Figure \ref{fig:iras2336all}. 
{\it Columns} (2) and (3) Peak position coordinates given with respect to $\alpha(J2000) = 23\h39\m01\fs0000$ and
$\delta(J2000) = 36\degr21\arcmin08\farcs000$. The errors in position (within parentheses), given in mas, were estimated 
by adding in quadrature the random errors in the position of the target (tar) and the phase reference calibrator (ref): 
$\sqrt[]{(\rmn{FWHM}_{\rmn{tar}}/(2\times SNR_{\rmn{tar}}))^2 + (\rmn{FWHM}_{\rmn{ref}}/(2\times SNR_{\rmn{ref}}))^2 }$,
where $SNR$ is the signal to noise ratio, and FWHM was taken as the projection of the beam 
major axis on both $\alpha$ and $\delta$ axes. {\it Column} (4) rms noise in the maps. {\it Column} (5) Peak intensities.  
{\it Column} (6) Flux densities measured in regions enclosing 5$\sigma$ level of the emission. {\it Column} (7) Matched 
C-band flux densities, covering the 5$\sigma$ level L-band emission region; this is done by adding the C-band $\stot$ to a 
3$\sigma$ emission in the remaining part. The uncertainties for the measurements shown in columns 5--7 
have been estimated by adding in quadrature the rms noise in the map plus a 5 per cent uncertainty in the point source calibration. 
{\it Column} (8) Characteristic size (in both $\alpha$ and $\delta$) occupied by the emission at a 5$\sigma$ level.)}} 
\label{tab:evnparam}
\begin{tabular}{@{}cccccccc@{}} \hline
Label & $\Delta\alpha$(J2000) & $\Delta\delta$(J2000) &        rms       & \speak{} & \stot{}
& $S_{\rmn{match}}$  &  $R_{\alpha}\times R_{\delta}$   \\
    & (\s{})  &  (\arcsec{})  & (\mujyb{})  & (\mujyb{})  & (\mjy{})   & (\mjy{})   & (pc$^2$)      \\
\multicolumn{1}{c}{(1)} & \multicolumn{1}{c}{(2)} & \multicolumn{1}{c}{(3)} & \multicolumn{1}{c}{(4)} & \multicolumn{1}{c}{(5)} 
   & \multicolumn{1}{c}{(6)} & \multicolumn{1}{c}{(7)} & \multicolumn{1}{c}{(8)} \\
\hline
L1 & 0.2600 (0.5)  & 0.592 (0.5)  &  28 & 786 $\pm$ 48  & 7.99 $\pm$ 0.40  &     $\cdots$    & 207 $\times$ 221  \\
C1 & 0.2614 (0.7)  & 0.603 (0.7)  &  16 & 303 $\pm$ 22  & 0.32 $\pm$ 0.02  & 1.33 $\pm$ 0.07 &  68 $\times$  69  \\
L2 & 0.2616 (0.7)  & 0.598 (0.7)  &  25 & 466 $\pm$ 34  & 5.42 $\pm$ 0.27  &     $\cdots$    & 221 $\times$ 220  \\
C2 & 0.2607 (0.5)  & 0.603 (0.5)  &  23 & 584 $\pm$ 37  & 1.11 $\pm$ 0.06  & 2.38 $\pm$ 0.12 &  98 $\times$  94  \\
L3 & 0.2615 (0.6)  & 0.566 (0.6)  &  30 & 640 $\pm$ 44  & 8.54 $\pm$ 0.43  &     $\cdots$    & 241 $\times$ 259  \\
C3 & 0.2608 (0.3)  & 0.599 (0.3)  &  18 & 875 $\pm$ 47  & 2.45 $\pm$ 0.13  & 3.52 $\pm$ 0.18 & 111 $\times$ 127  \\
    \hline
   \end{tabular} 
 \end{minipage}
\end{table*}

All the epochs were VLBI phase-referenced experiments using a data recording rate of 1024\,Mbps with two-bit 
sampling, for a total bandwidth of 128\,MHz. The telescope systems recorded both right-hand circular 
polarization (RCP) and left-hand circular polarization (LCP). The data were correlated at the EVN MkIV Data 
Processor at JIVE using an averaging time of 2\,s in the first two epochs, and 4\,s in the third one, since 
there was no need for a field of view (FOV) as large as 20\,arcsec. The sources  2134$+$004 and 3C45$4.3$ were used as 
fringe finders in all the observations. Each epoch lasted 6\,hr, from which a total of $\approx3.7$\,hr were 
spent on target. Target source scans of 3.5\,min were alternated with 1.5\,min scans of the phase reference source.

The correlated data of every epoch were analysed using the NRAO Astronomical Image Processing System (AIPS).  
The overall quality of the visibilities was good and the EVN pipeline products were useful for the initial steps
of the data reduction. To improve the calibration, we edited the data to remove artifacts due to radio interference 
(RFI) and included ionospheric corrections where needed. We exported the data of all the calibrators to the Caltech 
program DIFMAP \citep{difmap} and made images and visibility plots of each source. This allowed us to test the 
performance of each antenna and to determine gain corrections for each. When the gain correction was larger than 
10 per cent for a given antenna during the whole observing run, we applied it to the \uv{}data using the AIPS task CLCOR.

\begin{figure}
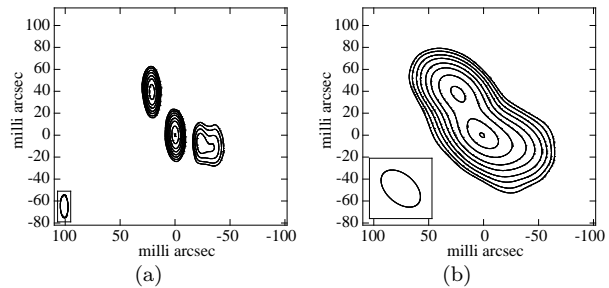

\centering
 \subfigure[]{
 \includegraphics[angle=0, scale=0.20]{j2333l1.eps} \label{fig:phcalL1}}
 \subfigure[]{
 \includegraphics[angle=0, scale=0.20]{j2333l2.eps} \label{fig:phcalL2}}
 \caption{EVN contour maps of the phase reference source J2333$+$3901 at L-band, first \subref{fig:phcalL1} and 
second epoch \subref{fig:phcalL2}.}\label{fig:phcal}         
\end{figure}

In the first two epochs we used J2333$+$3901 (at 2.9$\degr{}$ angular distance of target) as a phase reference source. 
This source has a complex structure (see Figure \ref{fig:phcal}) and it varied in flux density at L-band between
epochs (see column 7 in Table \ref{tab:evnobs}). The subtraction of the phase contribution due to the structure of 
J2333$+$3901 from the fringe solutions (delay and rate) was thus necessary. In spite of this correction, the phase 
referencing of the target source resulted in noisy phases. 

In the third epoch we used J2330$+$3348 (at 3.1$\degr{}$ angular distance of our target) as a phase calibrator, which being 
a predominantly compact source (at mas angular scales) provided a reliable phase reference and calibration. To correct the 
reference position in our first two epochs, and to align the three observing epochs, we obtained the shifts in right ascension 
and declination for the first two epochs that make their 15$\sigma$ emission coincide positionally with the 15$\sigma$ emission 
of the third epoch. We did this by means of the task UVSUB in AIPS, in which the data was divided by a point source model 
of 1\,Jy at the wanted reference position.

\subsection{Imaging process}\label{subsec:imag_evn}

The extended emission of IRAS\,23365 is not completely resolved with the available EVN array. The shortest baselines, such as 
Ef-Wb, can recover some of the extended emission. If no other short baselines are present (e.g. combinations of Jb, Cm and Kn), 
it is not possible to determine closure phases, and the presence of strong sidelobes (of the order of the peak) in the dirty map 
is thus favoured. This situation made it very difficult to obtain a reliable image of the target source \citep[see the preliminary 
maps of IRAS\,23365 presented in][]{rocc08}. In principle, removing such short baselines would solve the problem, at the expense 
of significantly degrading the final image sensitivity.

To properly map the extended emission, a good coverage of short baselines (resulting from combinations of at least three 
antennas to determine closure phases) is needed. To overcome the lack of short \uv{}spacings, a combination of Gaussian 
model fitting and imaging algorithms can be used. This is a widely used method for mapping the structure of outflows at VLBI 
scales \citep[see e.g.,][]{liza11}, specially for the cases in which faint diffuse emission is present together with the 
bright compact one. Epochs 1 and 2 were affected by poor short-baseline \uv{}coverage (the Cm-Kn baseline had a severe amplitude 
problem and it was not used). In epoch 3, baselines Jb-Kn, Jb-Cm, Cm-Kn and Ef-Wb were present, thus permitting to determine closure 
phases for the short baselines. As a result, no strong sidelobes affected the imaging process at this epoch. Nevertheless, for this 
epoch we also used a Gaussian model fitting combined with clean components in order to obtain consistent results with those of the 
first two epochs. This was done within the Caltech imaging programme DIFMAP \citep{difmap}. We exported the resulting images back into 
AIPS to analyse them and to produce the final maps that we present here (see Figure \ref{fig:iras2336all}). 

In Table \ref{tab:evnobs} we show the stations that participated in each observation. For different reasons, we 
lost some antennas and/or baselines and the final images were produced using the visibilities resulting from slightly different 
arrays. For instance, in the second epoch we lost Ur and Ar, and thus the resolution was compromised by the loss of the longest 
baselines. On the other hand, in the third epoch we had a good coverage of the short baselines (from combinations of Cm, 
Kn, Jb, Ef and Wb), which eased the reconstruction of the extended emission. To allow comparisons among the different epochs 
and frequencies, we used the same convolving beam (that from the epoch with the worst resolution: 26$\times$38\,mas$^2$ at 
46\degr{}) for the imaging process and sampled the beam using the same cell size ($=4$\,mas), and natural weighting for all 
epochs. The resulting images for the three EVN epochs at the two different frequencies (1.7 and 5\,GHz) are shown in
Figure \ref{fig:iras2336all}. The actual array used in the different epochs, is shown in a label at the upper right corner of 
each image. In Figure \ref{fig:combo_resol} (bottom) we also show the third EVN epoch at both 1.7 and 5\,GHz, as imaged with 
the natural beam of the observation at 1.7\,GHz (19$\times$25\,mas$^2$ at 30\degr{}).

\section{MERLIN and VLA observations}
\label{sec:comp_obs}

Simultaneously with our second EVN epoch, we also observed IRAS\,23365 at both L- and C-bands with MERLIN
(see Table \ref{tab:comboparam}), including the following antennas: Defford, Cambridge, Knockin, Darnhall, Mark 2 and 
Pickmere, observing with a bandwidth of 15\,MHz (in both circular polarisations). OQ208 was used as amplitude calibrator 
(1.1\,Jy at L-band and 2.5\,Jy at C-band) and J2333$+$3901 (0.8\,Jy at L-band and 0.34\,Jy at C-band) as phase calibrator. 
For phase-referencing, duty times of 7\,min/1.5\,min in L-band, and 2.5\,min/1.5\,min in C-band were used, for a total
time on source of 4 and 2.5\,hr at each band, respectively. 

We also analysed archival VLA (A-configuration) data at L- and C-bands \citep[project: AB660, reported in][]{baan06} 
to compare with the MERLIN and EVN images. The observations were performed with a bandwidth of 50\,MHz (in both circular 
polarisations). 3C48 (16.0\,Jy at L-band and 5.4\,Jy at C-band) was the flux calibrator and 0025$+$393 (0.7\,Jy and 0.6\,Jy at 
L- and C-band, respectively) the phase calibrator, which is found at 9.6\degr{} angular distance of the target source.

We followed standard procedures within AIPS for the data reduction. Details on the VLA and MERLIN observations are shown in 
Table \ref{tab:comboparam} and the resulting images in Figure \ref{fig:combo_resol}. We used matched baselines (in wavelengths)
to obtain the images for the two different frequencies of each array, to enable the comparison of information at the same scales. 
For the VLA images we used a common \uv{}range of 11.5 to 162.7 k$\lambda$ and same convolving beam of $1.3 \times 1.3$\,arcsec$^2$,
while we restricted MERLIN images to 112.6 to 1191.8 k$\lambda$, that resulted in a convolving beam of $0.10 \times 0.25$\,arcsec$^2$,
at 30$\degr{}$. We did not perform any \uv{}restriction in the case of the EVN data, to optimize the sensitivity and 
\uv{}coverage for each observing epoch. 

\begin{table*}
\centering
\begin{minipage}{168mm}
\caption{\protect{Parameters and observational data of complementary VLA and MERLIN observations
(see the corresponding contour maps in Figure \ref{fig:combo_resol}). The observations are labelled with 
a two-letter code, in which the first letter corresponds to the array (V$=$VLA or M$=$MERLIN), and the 
second letter to the observed frequency band ({\it Column} 3). {\it Columns} (4) and (5) Coordinates of 
the peak position, given with respect to $\alpha(J2000) = 23\h39\m01\fs0000$ and 
$\delta(J2000) = 36\degr21\arcmin08\farcs000$. {\it Column} (6) - rms noise in the maps. {\it Column} (7) Peak
intensities.  {\it Column} (8) Flux densities measured in regions enclosing 5$\sigma$ level of the emission.   
The uncertainties for the measurements shown in columns 4--5 and 7--8 have been estimated as described in Table
\ref{tab:evnparam}. {\it Column} (9) Major and minor axes, obtained by fitting a Gaussian to the source. 
{\it Column} (10) Characteristic size (in both $\alpha$ and $\delta$) occupied by the emission at a 
5$\sigma$ level.}}\label{tab:comboparam}
\begin{tabular}{@{}ccccccrrcc@{}} \hline
Label  & \multicolumn{1}{c}{Observing} &\multicolumn{1}{c}{Freq.}
& $\Delta\alpha$(J2000) & $\Delta\delta$(J2000) & \multicolumn{1}{c}{rms}  & \multicolumn{1}{c}{\speak{}} & 
 \multicolumn{1}{c}{\stot{}} & $\Theta_{\rmn{M}} \times \Theta_{\rmn{m}}$  & $R_{\alpha}\times R_{\delta}$   \\
  &    \multicolumn{1}{c}{date}  &  \multicolumn{1}{c}{(GHz)} & (\s{})  &   (\arcsec{})
& \multicolumn{1}{c}{(\mujyb{})} &  \multicolumn{1}{c}{(\mjyb{})}  &  \multicolumn{1}{c}{(\mjy{})} & (arcsec$^2$)  &  (kpc$^2$)    \\
\multicolumn{1}{c}{(1)} & \multicolumn{1}{c}{(2)} & \multicolumn{1}{c}{(3)} & \multicolumn{1}{c}{(4)} & \multicolumn{1}{c}{(5)} 
  & \multicolumn{1}{c}{(6)}  & \multicolumn{1}{c}{(7)}  & \multicolumn{1}{c}{(8)} & \multicolumn{1}{c}{(9)} & \multicolumn{1}{c}{(10)}\\
\hline
VL  & 1992-12-14 & 1.4  & 0.252 (6.1)& 0.54 (6.1) & 180 & 19.14 $\pm$ 0.97 & 25.18 $\pm$ 1.27 & 0.93 $\times$ 0.58 & 4.48 $\times$ 4.38\\
VC  & 1992-12-14 & 4.9  & 0.261 (3.3)& 0.59 (3.3) &  50 &  9.97 $\pm$ 0.21 & 10.82 $\pm$ 0.54 & 0.46 $\times$ 0.30 & 4.54 $\times$ 4.16\\
ML  & 2009-03-06 & 1.6  & 0.260 (1.6)& 0.55 (4.0) & 200 &  6.29 $\pm$ 0.37 & 13.90 $\pm$ 0.72 & 0.17 $\times$ 0.14 & 0.36 $\times$ 0.66\\
MC  & 2009-02-25 & 5.0  & 0.264 (3.6)& 0.56 (8.9) & 170 &  2.39 $\pm$ 0.50 &  5.20 $\pm$ 0.31 & 0.17 $\times$ 0.16 & 0.32 $\times$ 0.54\\
    \hline
   \end{tabular}
 \end{minipage}
\end{table*}
\begin{table*}
\centering
{\footnotesize
\begin{minipage}{180mm}
\caption{\protect{Physical quantities of IRAS\,23365 estimated from VLA, MERLIN and EVN observations. We have used the measured values of
\speak{}, \stot{} and the largest linear size between  $R_{\alpha}$ and $R_{\delta}$, from Tables \ref{tab:evnparam} and \ref{tab:comboparam}.
In the case of the EVN, we only include the third epoch of observations, re-imaged with a convolving beam of 19$\times$25\,mas$^2$ at 30\degr{}
(see Figure \ref{fig:combo_resol}). We consider the emission at both bands as 
measured from the region delimited by the 5$\sigma$ L-band emission, i.e., using \stot{} for L-band, and $S_{\rmn{match}}$ for C-band,
and the same linear size for both. {\it Column} (2) Brightness temperature calculated using as the source' solid angle 
$\Omega_{\rmn{s}}=\pi(4\rmn{log}2)^{-1}\left(\Theta_{\rmn{M}}\times\Theta_{\rmn{m}}\right)$ for the VLA and MERLIN (Table \ref{tab:comboparam});
for the EVN no reliable Gaussian fit can be made to the source (especially in the L-band map) in order to find its deconvolved size, and
we have used instead $\Omega_{\rmn{s}}=\pi(4\rmn{log}2)^{-1}\left(\rmn{FWHM}_{\rmn{M}}\times\rmn{FWHM}_{\rmn{m}}\right)$, 
where $\rmn{FWHM}_{\rmn{M}}$ and $\rmn{FWHM}_{\rmn{m}}$ are the major and minor synthesized beam fitted FWHM. {\it Column} (3) Monochromatic 
luminosity at the frequency $\nu$. {\it Column} (4) Two-point peak pixel-to-pixel spectral index determined from the pixel-to-pixel spectral
index distribution obtained with AIPS. {\it Column} (5) Two-point total spectral index determined between the L- and C-band flux 
densities ($\stot{}\propto\nu^{\alphtot}$). {\it Column} (6) Integrated isotropic radio luminosity, considering that the spectral behaviour of 
the source is straight between L- and C-band. {\it Column} (7) Equipartition magnetic field. {\it Column} (8) Characteristic lifetime of electrons 
subject to $B_{\rmn{eq}}$, undergoing radiative synchrotron losses.}}\label{tab:combovalues}
\begin{tabular}{@{}clcccccc@{}} \hline
\multicolumn{1}{c}{Label} & \multicolumn{1}{c}{log\,\tb{}} & \multicolumn{1}{c}{$L_{\nu}$} &\multicolumn{1}{c}{\alphpp{}} &
 \multicolumn{1}{c}{\alphtot{}} &  \multicolumn{1}{c}{$L_{\rmn{R}}$} &  \multicolumn{1}{c}{$B_{\rmn{eq}}$} &  
   \multicolumn{1}{c}{$\tau_{\rmn{syn}}$} \\
 & \multicolumn{1}{c}{(K)} & \multicolumn{1}{c}{($10^{29}$\lunits{})} &  &
   &  \multicolumn{1}{c}{($10^{39}$\,\rlunits{})} &  \multicolumn{1}{c}{($\mu$G)}  
    &\multicolumn{1}{c}{(Myr)}  \\ 
\multicolumn{1}{c}{(1)} & \multicolumn{1}{c}{(2)} & \multicolumn{1}{c}{(3)} & \multicolumn{1}{c}{(4)} & \multicolumn{1}{c}{(5)} 
   & \multicolumn{1}{c}{(6)} & \multicolumn{1}{c}{(7)} & \multicolumn{1}{c}{(8)}  \\
\hline
VL & 4.45 $\pm$ 0.02  & 19.11 $\pm$ 0.97 &                                   &                                   &                      & 18.4 & 11.4 \\[-1ex]
VC & 3.61 $\pm$ 0.02  &  8.21 $\pm$ 0.41 &\raisebox{1.5ex}{-0.46 $\pm$ 0.06} &\raisebox{1.5ex}{-0.69 $\pm$ 0.06} &\raisebox{1.5ex}{4.1} & 18.1 &  6.2 \\
ML & 5.42 $\pm$ 0.02  & 10.55 $\pm$ 0.55 &                                   &                                   &                      & 77.1 & 1.1  \\[-1ex]
MC & 3.96 $\pm$ 0.03  &  3.95 $\pm$ 0.24 &\raisebox{1.5ex}{-0.66 $\pm$ 0.13} &\raisebox{1.5ex}{-0.89 $\pm$ 0.07} &\raisebox{1.5ex}{2.0} & 90.8 & 0.6  \\
EL & 6.81 $\pm$ 0.02  &  4.94 $\pm$ 0.25 &                                   &                                   &                      & 174.7&  0.4 \\[-1ex]
EC & 5.57 $\pm$ 0.02  &  2.58 $\pm$ 0.13 &\raisebox{1.5ex}{ 0.52 $\pm$ 0.13} &\raisebox{1.5ex}{-0.59 $\pm$ 0.06} &\raisebox{1.5ex}{1.1} & 174.7&  0.2 \\
    \hline
   \end{tabular}
 \end{minipage}}
\end{table*}

\section{Results}

IRAS\,23365 has been observed at different resolutions (EVN, MERLIN and VLA) and at two different frequencies. This allows a
comparison among the different linear scales mapped with different arrays. In the following, we present our results regarding
morphology, radio emission, radio spectrum and magnetic field of IRAS\,23365. The different parameters measured from the three
epochs of EVN observations (see Figure \ref{fig:iras2336all}) are presented in Table \ref{tab:evnparam}, and measurements 
from the VLA archival data and the MERLIN observations are presented in Table \ref{tab:comboparam}. In Table \ref{tab:combovalues}
we show the estimates from the measurements at different scales as shown in Figure \ref{fig:combo_resol}. 

\subsection{The IRAS\,23365 structure: from kpc- down to pc-scales}\label{sec:obs_str}

The radio images of IRAS\,23365 shown in Figure \ref{fig:combo_resol}, cover the structure of this galaxy at three
different scales: galactic (with the VLA), circumnuclear (with MERLIN) and nuclear (with the EVN). We use 
the third epoch of EVN observations to compare with the VLA and MERLIN images, since that was the epoch with the 
best compromise between angular resolution and sensitivity (see Section \ref{subsec:imag_evn} for details). For doing this,
we re-imaged the third EVN epoch (L- and C-band maps) using as convolving beam that one which resulted from the L-band 
(19$\times$25\,mas$^2$ at 30\degr{}, see bottom of Figure \ref{fig:combo_resol}).

At galactic scales, the emission at both L- and C-bands is unresolved and appears concentrated in a zone of $\sim 4.5$\,kpc
in size. At circumnuclear scales, the emission is concentrated in the inner 0.5 kpc region and displays some extended structure 
on top of an unresolved component.

At the highest resolution in L-band, the nuclear region has a size $\ga$200\,pc at all the EVN epochs (see Table \ref{tab:evnparam}), 
and shows variations in its morphology (see Figure \ref{fig:iras2336all}). At C-band, the emitting region is about 100\,pc and its 
structure remains quite compact in the first two epochs, whilst some more extended emission is traced in the third  
epoch (see Figure \ref{fig:iras2336all}). A single Gaussian fit is inaccurate for obtaining the deconvolved size of the emitting region, 
at least for the emission at L-band, due to the wealth of extended emission. We thus characterize the area of the emitting region 
with the size of the source in both the right ascension and declination axes, $R_{\alpha}$ and $R_{\delta}$, respectively (Table 
\ref{tab:evnparam}). There are some features outside the nuclear regions that, while having peaks slightly above $5\sigma$ in our 
second epoch (Figures \ref{fig:L2} and \ref{fig:C2}), are not seen neither in our first epoch, nor in our third observing epoch. 
While these could be real features (in particular, the compact source detected at both frequencies with $\Delta \alpha$, 
$\Delta \delta \sim 150$\,mas), we conservatively consider them as tentative detections (see Section \ref{subsec:imag_evn} for details) 
and therefore are not discussed here.

We note that the size of the emission area increases with time through the different EVN epochs at both frequencies (see Table 
\ref{tab:evnparam}), and also displays different morphology (see Figure \ref{fig:iras2336all}), especially at L-band. Whereas 
sensitivity does not seem to vary drastically among epochs, the observations were performed at different hour angles and thus the 
\uv{}plane was sampled at different orientations. Hence, the differences in size and morphology could have been affected by the 
different \uv{}coverages. 

Regardless of the used array (i.e., VLA, MERLIN or EVN), and albeit of using matched baselines (at least for VLA and MERLIN),
the emission in L-band consistently occupies a larger extension than that at C-band, around a factor of 2 in the case of the EVN, 
as seen in Figure \ref{fig:combo_resol}, where we show for comparison the VLA, MERLIN and EVN (third epoch) images. This can be 
explained by the longer lifetime of accelerated electrons emitting synchrotron radiation at lower frequencies (see Section 
\ref{sec:magfield}).

We also note that the peak positions at the two different frequencies are not coincident neither for MERLIN nor for the EVN. In 
the case of the VLA, we do not have the required angular resolution to confirm any shift; however, at the higher resolution provided 
by both MERLIN and EVN, a shift of the C-band peak towards the North-East direction is evident, while that at L-band is shifted 
towards the South-West (see Tables \ref{tab:evnparam} and \ref{tab:comboparam} and Figure \ref{fig:combo_resol}). This result is
consistent for all the epochs and at the different angular resolutions provided by EVN and MERLIN. We interpret those 
shifts of the emission peaks as evidence for at least two different populations of radio emitters being present in the nuclear 
region (see Section \ref{subsec:nucspec}). Furthermore, the peak component is variable both in position and in intensity among 
EVN epochs, and in each epoch, being also different between frequencies. These facts give evidence of the source variability 
within the innermost nuclear region.

\subsection{The radio emission and radio spectrum at different scales}\label{subsec:nucspec}

We mentioned in the previous section that the radio emission at different frequencies seen at the different resolutions (except 
perhaps for the VLA), peaks at different positions. Thus, a peak spectral index defined as $\speak{}\propto \nu^{\alphpeak}$ 
would be meaningless. We use instead the peak of the pixel-to-pixel spectral index distribution (\alphpp{}) as obtained with AIPS.

The radio emission of both galactic and circumnuclear regions of IRAS\,23365 mapped with the VLA and MERLIN, respectively, 
is stronger at L-band than at C-band (see columns 7 and 8 in Table \ref{tab:comboparam}). Consequently both total spectral 
indices ($\stot{}\propto \nu^{\alphtot}$) and peak pixel-to-pixel spectral indices (\alphpp{}) are steep, as shown in columns 
4 and 5 of Table \ref{tab:combovalues}. Steep spectral indices are an indication of non-thermal emission. We note however that 
the \tb{} values (column 2 in Table \ref{tab:combovalues}) at galactic (L- and C-bands) and circumnuclear (C-band) scales, are 
in principle consistent with either thermal emission, or with synchrotron emission suppressed by free-free absorption from e.g., 
\hii{} regions. The calculated value for the free-free opacity ($\tau_{\rmn{ff}}$) implies that thermal emission should be optically 
thick. Therefore, the bulk of emission at L- and C-bands corresponds to optically thin non-thermal synchrotron emission, slightly 
affected by free-free absorption at galactic ($\tau_{\rmn{ff}}\approx0.02$) and circumnuclear ($\tau_{\rmn{ff}}\approx0.20$) scales.

Regarding the nuclear region (mapped with the EVN), the large \tb{} values are consistent with pure non-thermal emission. The attained 
angular resolution and the presence of strong extended (a few mJy; see Tables \ref{tab:evnparam}, and \ref{tab:comboparam}) radio
emission, prevents us from directly detecting individual faint (see Table \ref{tab:evnsample}) compact sources, e.g. SNe. However, 
we note that \speak{} and \stot{} show variations at both frequencies during our EVN monitoring campaign (see Table \ref{tab:evnparam}).
Whereas in C-band, \speak{} and \stot{} increase with time, in L-band these diminish in the second epoch, and then increase in the 
third one, thus indicating the variability of sources and/or the appearance of new ones within the nucleus, e.g., new SNe, accounting 
to the expected SN rate ($\approx2.4$\,yr$^{-1}$). This non-correlated behaviour at both frequencies is indicative of nuclear activity
that becomes transparent first at C-band and later at L-band.

\subsection{Spectral index distribution at mas-scales}\label{subsec:specmas}

Let us now concentrate in the spectral indices corresponding to the EVN images. Considering the total flux densities as measured 
from the region within the 5$\sigma$ L-band emission (i.e. L-band flux from column 6, and C-band flux from column 7 in Table 
\ref{tab:evnparam}), the total spectral indices (\alphtot{}) are steep for all the epochs. However, the situation is different
for the peak pixel-to-pixel spectral index (\alphpp{}), which is evolving with time. The distribution of  \alphpp{} is shown in 
Figure \ref{fig:alphapeak} for the three EVN epochs. In the first epoch of EVN observations, \alphpp{} is steep, then it becomes 
inverted in our second epoch, and starts to decrease (although being still inverted) in the third epoch to presumably become steep 
again. This is clear evidence of the variation in flux of sources within the innermost nuclear regions, and/or appearance of new 
sources (e.g., SNe) which would be seen first at higher frequencies and later on at lower frequencies \citep{weiler02}, in agreement 
with our results. We also note that for the three EVN epochs, \alphpp{} (which is given pixel by pixel as shown in Figure 
\ref{fig:alphapeak}), becomes steeper as measured towards the edges of the C-band emission, where the noise at C-band starts to 
dominate, whilst there is still extended emission detected at L-band. This is a consequence of the ageing of the population of 
electrons radiating synchrotron emission (see Section \ref{sec:magfield}).

\subsection{The magnetic field in the energy budget of IRAS\,23365}\label{sec:magfield}

In previous sections, we have gathered information about the ongoing non-thermal activity of IRAS\,23365 at different scales. 
In a ULIRG environment, we expect SNe, SNRs and/or an AGN to be the engines responsible for producing high energy particles which 
will interact with the galactic magnetic field, thus generating synchrotron radiation \citep[dominating at $\nu \la 30$\,GHz;][]{condon}. 
The energy thus produced, will be present in the form of relativistic particles and magnetic field. In the following, we investigate 
the energy budget (due to synchrotron radiation) of IRAS\,23365 at different scales, i.e., as estimated from observations with different 
arrays (EVN, MERLIN and VLA). We only consider the third epoch of observations with the EVN, to compare with 
the results from the VLA and MERLIN, since this epoch was the one which had the least imaging problems (see Section 
\ref{subsec:imag_evn}).

We can estimate the average equipartition magnetic field based on the radio emission of IRAS\,23365 as follows 
\citep[see][]{pachol},

{\scriptsize
\begin{equation}\label{eq:Beq}
\left(\frac{B_{\rmn{eq}}}{\mu\rmn{G}}\right) \approx 8.1 \left[ \frac{(1+k)}{\phi} \left(\frac{c_{\mbox{\tiny 12}}}{10^7}\right)
            \left(\frac{R}{1\,\rmn{kpc}} \right)^{-3} \left(\frac{ L_{\rmn{R}}}{10^{39}\,\rlunits{}}\right) \right]^{2/7}
\end{equation}}
\noindent
where $\phi$ is the filling factor of fields and particles, $k$ is the ratio of heavy particle energy to electron energy,
and $c_{\mbox{\tiny 12}}$ is a function that depends on the minimum and maximum frequencies considered, and of the two-point spectral 
index, $\alphtot$, which is estimated based on those two frequencies \citep[see ][]{pachol}. $L_{\rmn{R}}$ is the integrated 
isotropic radio luminosity between the minimum and maximum frequencies used, and $R$ is the linear size occupied by the emission, 
taken as the larger value between $R_{\alpha}$ and $R_{\delta}$ in each case (see column 10 in Table \ref{tab:comboparam} for the VLA 
and MERLIN). For the third EVN epoch, we determined a maximum linear size $R\approx0.2$\,kpc with TVDIST within AIPS.
For simplicity, we consider $\phi=0.5$ and $k=100$ \citep[see e.g.,][]{miguel07}. 

In Table \ref{tab:combovalues} we show the average values for \alphtot{}, $L_{\rmn{R}}$ and $B_{\rmn{eq}}$, obtained within the 
emission regions sampled by the different instruments. In the innermost nuclear region (imaged with the EVN), the strength of the 
magnetic field is larger than the one measured at lower resolutions. This is expected, since the plasma in the central regions should 
be denser than in the outer regions, and thus the magnetic field lines therein, frozen within the plasma, should be more concentrated. 
The average magnetic field under energy equipartition, for the emission measured with the VLA, MERLIN and the EVN, would be 18, 
84 and 175 $\mu$G respectively. The latter value represents the peak of $B_{\rmn{eq}}$ coming from the very central region. If the 
synchrotron spectrum holds beyond the C-band frequencies, e.g. to 20\,GHz, the estimated values for $B_{\rmn{eq}}$, would only be 
$\sim15$ per cent larger.

Our obtained $B_{\rmn{eq}}$ value at galactic scales is consistent with that of a galaxy in advanced interaction state, probably close 
to nuclear coalescence, according to VLA studies of interacting galaxies by \citet{drzazga11}. Likewise, the $B_{\rmn{eq}}$ value
at nuclear scales is similar to that found through VLBI studies of the ULIRG IRAS\,17208-0014 \citep[144\,$\mu$G;][]{momjian03}, which 
is also an advanced merger.

Considering the obtained values for $B_{\rmn{eq}}$, and following \citet{pachol}, we can calculate the lifetime of the electrons 
with energy $E_\rmn{min}$, which move in a magnetic field of strength $B_{\rmn{eq}}$, thus emitting synchrotron radiation around 
a critical frequency $\nu_{\rmn{c}}$. This is,
\[ \tau_{\rmn{syn}} = \frac{1}{c_2 E_\rmn{min} B_{\rmn{eq}}^2}, \]
(where $c_{\mbox{\tiny 2}}$ is a constant)
\begin{equation}\label{eq:tsyn}
\Rightarrow \left( \frac{\tau_{\rmn{syn}}}{\rmn{Myr}}\right) \approx
1.06\times 10^3\left[\left(\frac{\nu_{\rmn{c}}}{\rmn{GHz}}\right)\left(\frac{B_{\rmn{eq}}}{\mu \rmn{G}}\right)^3\right]^{-1/2}
\end{equation}

\begin{table*}
\centering
\begin{minipage}{168mm}
\caption{\protect{Parameters derived for the compact components EC1, EL1 and EL2 found within the extended emission in the third epoch
of EVN observations (Figure \ref{fig:combo_resol}). {\it Columns} (2) and (3) Coordinates of the peak position, given with respect to 
$\alpha(J2000) = 23\h39\m01\fs0000$ and $\delta(J2000) = 36\degr21\arcmin08\farcs000$. {\it Column} (4) Peak intensity corrected by the zero 
level emission. {\it Column} (5) Monochromatic luminosity. {\it Column} (6) Brightness temperature, considering
that the solid angle subtended by the source is that subtended by the synthesized beam (as in Table \ref{tab:combovalues}).}}\label{tab:cc_l3}
\begin{tabular}{@{}cccccc@{}} \hline
Label & $\Delta\alpha$(J2000) & $\Delta\delta$(J2000) &  \speak{} &  \multicolumn{1}{c}{$L_{\nu}$}  &  \multicolumn{1}{c}{log\,\tb{}} \\
    &  (\s{})  &  (\arcsec{})  &    \multicolumn{1}{c}{(\mujyb{})} & \multicolumn{1}{c}{($10^{28}$\lunits{})}  &\multicolumn{1}{c}{(K)} \\
\multicolumn{1}{c}{(1)} & \multicolumn{1}{c}{(2)} & \multicolumn{1}{c}{(3)} & \multicolumn{1}{c}{(4)} & \multicolumn{1}{c}{(5)} 
   & \multicolumn{1}{c}{(6)} \\
\hline
EC1 &  0.2608 (0.4)  &  0.599 (0.4)   &  352 $\pm$ 35  &  2.67 $\pm$ 0.26 &  4.56 $\pm$ 0.04  \\
EL1 &  0.2597 (1.7)  &  0.586 (1.7)   &  150 $\pm$ 32  &  1.14 $\pm$ 0.24 &  5.15 $\pm$ 0.09  \\
EL2 &  0.2616 (1.4)  &  0.561 (1.4)   &  184 $\pm$ 33  &  1.39 $\pm$ 0.25 &  5.23 $\pm$ 0.08  \\
    \hline
   \end{tabular}
 \end{minipage}
\end{table*}

From Section \ref{sec:obs_str}, we know that the radio emission at L- and C-bands has a different extent and peaks at 
different positions within the nuclear region, which strongly suggests the presence of different populations of particles. 
This is more evident in the nuclear region mapped with the EVN: in the innermost region, where there is an overlap 
between the emission at the two different frequencies, there would be a concentration of very energetic, short-lived particles, 
whereas the outer region would be populated by less energetic, long-lived particles, which had had time to diffuse from the 
inner regions into the outer ones. In Table \ref{tab:combovalues} we show the values for $\tau_{\rmn{syn}}$, assuming that the 
critical frequency is either that of the L-band or the C-band. In all cases we see that the L-band emission is tracing the 
emission from an older population of electrons (regardless of the resolution) than the one emitting at C-band frequencies. 
The putative AGN together with an ensemble of SNe, for which evidence has been found in other studies (see Section 
\ref{sec:iras2336}), must be located within the C-band emission region as seen with the EVN, where the magnetic field strength 
is larger, and where a composite spectrum (which varies with time) has been found (Section \ref{subsec:nucspec}).

We note that the radio lifetime of the emitting source is not only determined by $\tau_{\rmn{syn}}$. The lifetime of relativistic
electrons might also be affected by Compton losses given by
\[ \tau_{\rmn{C}} = \frac{25.2}{U_{\rmn{rad}}E_{\rmn{min}}}, \]
\citep[following][]{pachol} since the electrons are immersed in a radiation field,
\[U_{\rmn{rad}}= \frac{4\pi}{c}\frac{L_{\rmn{bol}}}{\Omega_{\rmn{s}}},\]
for which we take \lir{} as a good approximation to the bolometric luminosity $L_{\rmn{bol}}$.
$U_{\rmn{rad}}$ varies from $\approx 2.5\times 10^{-7}$\,erg\,cm$^{-3}$ at galactic scales,
up to $\approx 2.8\times 10^{-4}$\,erg\,cm$^{-3}$ at nuclear scales. 
To compare the different losses, we calculate their ratio,
\[ \tau_{\rmn{C}}/\tau_{\rmn{syn}} = \frac{25.2 c_2 B_{\rmn{eq}}^2}{U_{\rmn{rad}}}.\]
We note that at all scales (nuclear, circumncuclear and galactic) and at both L- and C-band, we obtain
$\tau_{\rmn{syn}} \gg \tau_{\rmn{C}}$, with a ratio ranging between $6.5\times 10^{-6}$ (nuclear scales) and
$8.8\times 10^{-5}$ (galactic scales); i.e., the energy density of the radiation field, greatly exceeds the magnetic 
energy density. Radiative losses are important and we argue that there is need for injection of new electrons or a 
continuous acceleration to halt the energy depletion, otherwise radio emission would not be visible.

The re-acceleration or injection of new electrons in a (U)LIRG environment, is very likely provided in SN-shells and SNRs
by first order Fermi acceleration. The presence of SNe, SNRs and a strong magnetic field in IRAS\,23365, agrees with this 
scenario.

\subsection{The nuclear region in the third EVN epoch}
\label{sec:compactEL}

Among the EVN observing epochs, the third one benefited from a better \uv{}coverage, and thus resulted in a smaller natural 
beam (19$\times$25\,mas$^2$ at 30\degr{} at L-band). The L-band map (Figure \ref{fig:combo_resol}, bottom-left) shows the presence of two
compact sources (EL1 and EL2) within the nuclear region, without counterparts at C-band. On the other hand, the compact source 
that dominates the emission at C-band, labelled as EC1 (Figure \ref{fig:combo_resol}, bottom-right), has no compact counterpart 
at L-band, although extended emission is present.

To obtain EC1, EL1 and EL2 peak intensities, we first estimated the background emission where 
these compact sources lay. We solved for the 'zero level' emission ($S_{\rmn{bg}}\sim 179\pm24$\,\mujy{} at C-band and 
$S_{\rmn{bg}}\sim 211\pm12$\,\mujy{} at L-band) using the task IMFIT within AIPS. We then subtracted this value from 
the maximum intensity found at the positions of each compact source, in order to obtain their \speak{}. In Table \ref{tab:cc_l3} 
we give the positions for EC1, EL1 and EL2, their peak intensities, as well as their estimated $L_{\nu}$ and \tb{}, which are 
indicative of a non-thermal origin. 

EC1 lays on a region where \alphpp{} changes with time, suggesting variability within this region. We note that EC1 is confined 
to a small area in the first epoch, and then appears to increase in size, as we have mentioned in Section \ref{sec:obs_str}. 
EL1 lays in a region with basically no C-band emission, unlike the region where EL2 lays where more extended 
emission is being traced from the first epoch to the third one at C-band. As a consequence, EL1 lays in a region which maintains
a very steep \alphpp{} through time, whilst EL2 is found in a region with varying \alphpp{} (Figure \ref{fig:alphapeak}). 
However, we cannot rule out that the \alphpp{} variations at EL2 are intrinsic.

In the case of EC1, both the variability of the radio emission (see Section \ref{subsec:nucspec}) and of the spectral index 
distribution (see Section \ref{subsec:specmas} and Figure \ref{fig:alphapeak}) are indicative of recent non-thermal activity 
(probably due to SNe and/or AGN activity). EL1 and EL2 display brightness temperatures similar to those expected from either type 
II SNe or SNRs. We note that the maximum linear size for EL1 and EL2 is set by the beam size to $\sim 30$\,pc, which is too large 
for characterising either an individual SN or a SNR. A scenario in which EL1 and EL2 are clusters of SNe is difficult to 
reconcile with the absence of peaks of emission at C-band in all the EVN observing epochs, and with the behaviour of \alphpp{}
at both EL1 and EL2. These facts suggest that there is no recent activity from young SNe in those regions, and favour
an scenario in which EL1 and EL2 are dominated by an old population of radio emitters.

\section{Summary and discussion}

We have presented state-of-the-art radio interferometric images of IRAS\,23365, one of the brightest and farthest ULIRGs
in the local Universe ($z<0.1$). 

Our images reveal the presence of a nuclear region, possibly a starburst-AGN composite, with an approximate size of 200\,pc in L-band, 
and about 100\,pc in C-band. We find that the L- and C-band radio emission peak at different positions, thus suggesting that the nuclear 
region is composed of at least two zones, dominated by distinct populations of radio emitters. 

In the region where the L- and C-band emission overlap, there is evidence for ongoing non-thermal activity, characterised by very 
energetic, short-lived particles . During our EVN monitoring of IRAS\,23365, we have found flux density variability in the overlapping 
region, thus resulting in a variation of the spectral index. This can be explained by the flux density variations of sources therein 
(SNe, AGN, etc.) and/or by the appearance of new sources (e.g., SNe) which would be seen first at higher frequencies and later at lower 
frequencies \citep{weiler02}. The edges of the overlapping region characterised by less energetic, long-lived particles, would be 
dominated by an old population of radio emitters, probably clumps of SNRs, for which we have found two candidates in the third L-band
EVN epoch. These facts agree with the classification of IRAS\,23365 as a composite system, made by \citet{yuan10}.

The radio source lifetime at different scales (as seen with the VLA, MERLIN and the EVN arrays) and at both L- and C-bands,
is limited by Compton losses. The SNe and SNRs, for which we have found evidence, are likely providing the
mechanism of re-acceleration, or replenishment of new electrons that is needed to halt the radio energy depletion. 

We have found that the equivalent magnetic field strength at galactic (mapped with the VLA) and nuclear scales (mapped with the EVN),
18 and 175\,$\mu$G, respectively, correspond to that of a galaxy in an advanced stage of interaction \citep{drzazga11,momjian03}. 
The magnetic field in both nuclear and circumnuclear regions is stronger than at galactic scales, thus implying that the lifetime of 
the electrons undergoing synchrotron losses is shorter ($\la1$\,Myr) in the innermost nuclear regions (with linear size $R\la0.5$\,kpc) 
of IRAS\,23365, and larger ($\ga1$\,Myr) in the outer regions ($R\ga4$\,kpc). 

Our study of IRAS\,23365 (at $z \sim 0.06$) has shown that high-resolution, high-sensitivity observations are needed if we are to 
make significant improvement in the detailed understanding of nuclear and circumnuclear starbursts in the local Universe. 
The resolution we attained using a maximum baseline length of approximately 7,000\,km, is not enough to resolve individual
compact sources (e.g., SNe, SNRs, AGN) from each other, within the nuclear region of IRAS\,23365; yet, it could be possible to infer
the activity of such compact sources, by carefully monitoring variations of total flux density and spectral index distribution.
For instance, \citet{rocc11a} were able to directly detect SN activity in the B1-nucleus of Arp\,299, by carefully monitoring the 
variations in \stot{}, over several years of VLA observations. Without spatially resolving each individual SNe, they estimated a lower 
limit for the \snrate{} in that LIRG. In the case of IRAS\,23365, where a larger number of SNe are expected each year, several 
observations per year would be needed to perform such an indirect study of the SN population in its nuclear region, provided that we 
are able to distinguish between AGN outbursts and SN explosions. 

IRAS\,23365 is a good example of the situation to be faced when observing galaxies at higher redshifts. It is expected that the 
Square Kilometer Array (SKA), with a maximum baseline length $\approx3,000$\,km, will allow the detection of sources as faint as 
50\,nJy, e.g. CCSNe, exploding at $z\sim5$ \citep[see e.g.,][]{lien11}. However, the angular resolution will be a strong limiting 
factor. In those cases where the nuclear and even the circumnuclear regions (i.e., where we expect most of the SN activity to occur) 
of the host galaxy cannot be resolved out into their different components, SKA's high sensitivity might be of great use to indirectly 
detect SN activity through the monitoring of flux density variations.

\begin{figure*}
\centering
\subfigure[]{
 \includegraphics[angle=0, scale=0.28]{iras2336l1.eps} \label{fig:L1}}
 \subfigure[]{
 \includegraphics[angle=0, scale=0.28]{iras2336c1.eps} \label{fig:C1}}
 \subfigure[]{
 \includegraphics[angle=0, scale=0.28]{ir2336_cl1.ps} \label{fig:CL1}}
\subfigure[]{
 \includegraphics[angle=0, scale=0.28]{iras2336l2.eps} \label{fig:L2}}
 \subfigure[]{
 \includegraphics[angle=0, scale=0.28]{iras2336c2.eps} \label{fig:C2}}
 \subfigure[]{
 \includegraphics[angle=0, scale=0.28]{ir2336_cl2.ps} \label{fig:CL2}}
\subfigure[]{
 \includegraphics[angle=0, scale=0.28]{iras2336l3.eps} \label{fig:L3}}
 \subfigure[]{
 \includegraphics[angle=0, scale=0.28]{iras2336c3.eps} \label{fig:C3}}
 \subfigure[]{
 \includegraphics[angle=0, scale=0.28]{ir2336_cl3.ps} \label{fig:CL3}}
 \caption{IRAS\,23365 at L-band (left: \subref{fig:L1}, \subref{fig:L2}, \subref{fig:L3}), 
   C-band (middle: \subref{fig:C1}, \subref{fig:C2}, \subref{fig:C3}), and L-band contours 
   overlaid on grey scale C-band images (right: \subref{fig:CL1}, \subref{fig:CL2}, \subref{fig:CL3}), 
   in three different epochs (top to bottom), with the grey scale in \mujy{} ranging from 3$\sigma$
   value of the noisiest map in C-band (second epoch, i.e., $\sigma=23\mujyb{}$), up to the peak intensity value of each epoch
   at C-band. Dashed contours represent $-3\sigma$ levels. All the images have been degraded to the epoch with lowest 
   resolution (L2, following the labels in Table \ref{tab:evnobs}), being therefore convolved with the same beam 
   size: 26$\times$38\,mas$^2$ at 46\degr{}. The maps are centred at 23\h39\m01\fs29, +36\degr21\arcmin08\farcs59  
   (J2000). At the top of each image, we provide the EVN project code, observation date, observed 
   wavelength and the list of antennas used for mapping. In the lower left corner we show the peak intensity, noise
   and contour levels information. The size of the nuclear region seems to increase from one epoch to another at
   both frequencies, although images from the same frequency were convolved with the same beam. We consider that this
   is an effect of the difference in the arrays used, rather than being an intrinsic change in the source, albeit
   this possibility cannot be neglected. It is worth noting that the size of the nuclear zone is consistently larger
   at L-band than its counterpart at C-band. This can be explained by the longer lifetime at lower frequencies of the 
   electrons being accelerated in the innermost nuclear regions.}\label{fig:iras2336all}         
\end{figure*}

\begin{figure*}
\centering
\includegraphics[scale=0.65]{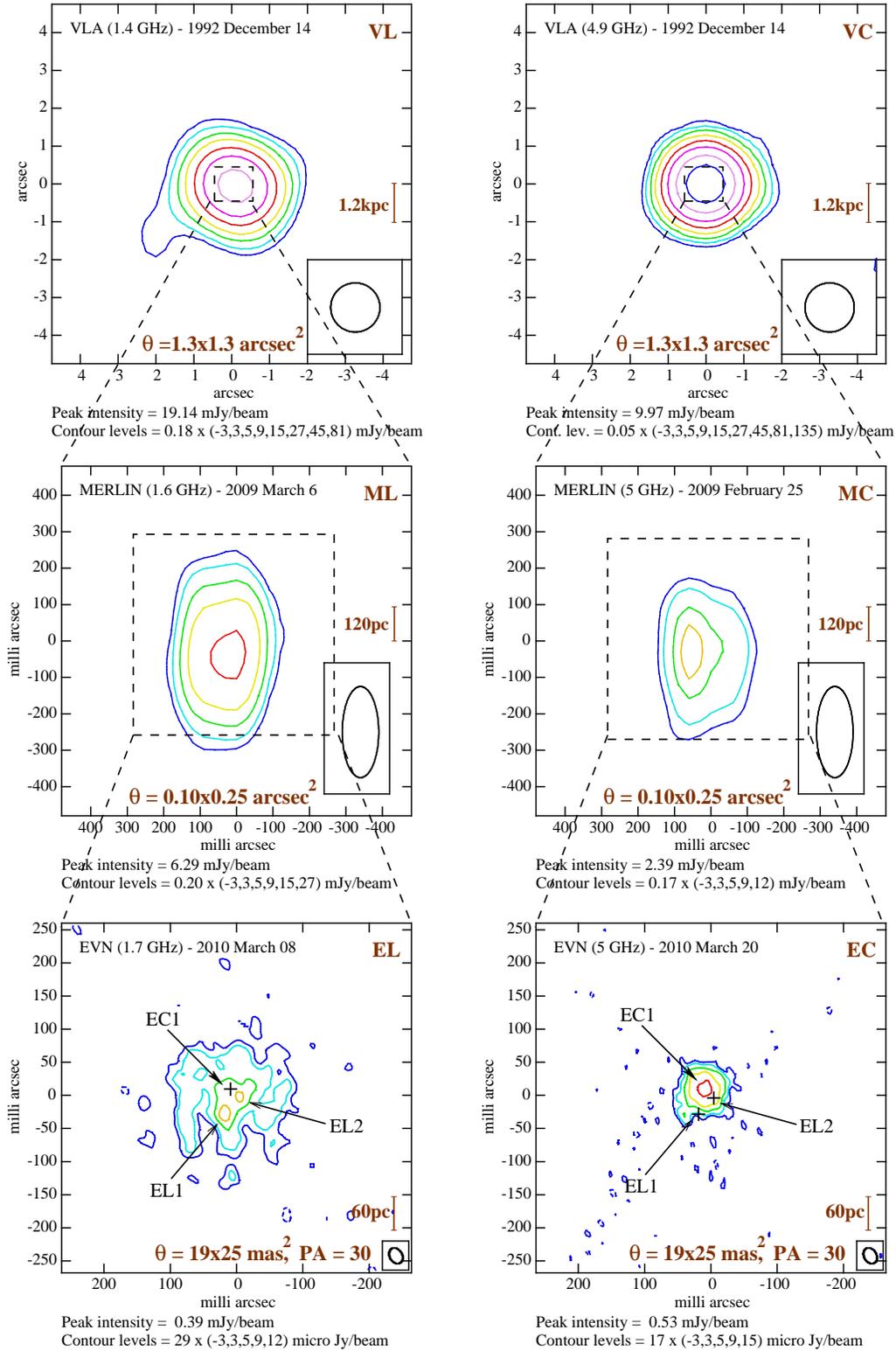}
\caption{IRAS\,23365 as seen at different frequencies and resolutions (higher resolution from top to bottom; and higher 
frequency from left to right, see labels). All the maps are centred at 23\h39\m01\fs29, +36\degr21\arcmin08\farcs59  
(J2000). Dashed contours represent $-3\sigma$ levels. The cross in the EL image indicates the peak position (EC1) 
as measured from the EC image, and the crosses in the 
EC image indicate the positions of components EL1 and EL2. The non-coincidence of the peaks at the two different frequencies 
(see also ML and MC images), indicates the presence of distinct population of sources in the innermost nuclear region. 
We note that at C-band, IRAS\,23365 remains practically unresolved, whilst L-band images show more extended structure.} 
\label{fig:combo_resol}
\end{figure*}

\begin{figure}
\centering
\includegraphics[scale=0.68]{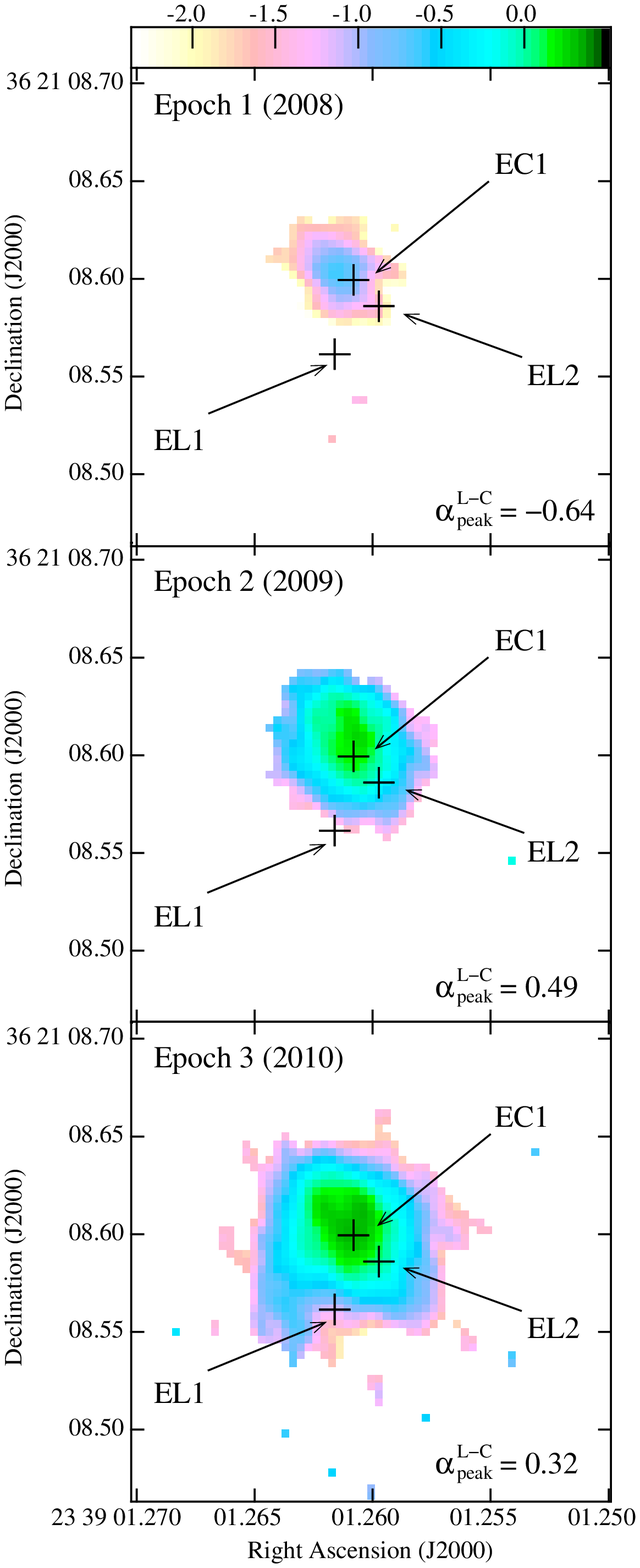}
\caption{Evolution of the spectral index distribution ($S_{\rmn{pixel}} \sim\nu^{\alphpp}$), given pixel by pixel (as 
obtained from AIPS) in the nuclear zone, which we expect to be composed by non-thermal sources (e.g. SNe and AGN). Note 
that while the peak of the spectral index distribution varies in the very central region, it becomes steeper towards the 
edges, thus representing older radio emission. The sources EC1, EL1 and EL2 are marked with crosses (see Section 
\ref{sec:compactEL}).}\label{fig:alphapeak}\end{figure}

\clearpage

\section*{Acknowledgements}
We thank our referee Robert Beswick for constructive comments and suggestions that have improved this manuscript.
We are grateful to the editor for useful suggestions on improving the presentation of our results.
We acknowledge financial support from the Spanish MICINN through grant AYA2009-13036-C02-01, co-funded with FEDER
funds. We also acknowledge support from the Autonomic Government of Andalusia under grants P08-TIC-4075 and TIC-126. 
Our work has also benefited from research funding from the  European Community Framework Programme 7, 
Advanced Radio Astronomy in Europe, grant agreement no.: 227290, and sixth Framework Programme under RadioNet R113CT 
2003 5058187. The authors are grateful to JIVE and especially to Zsolt Paragi and Bob Campbell for their assistance in 
this project. \emph{The European VLBI Network is a joint facility of European, Chinese, South African and other radio 
astronomy institutes funded by their national research councils}. This article is also based on observations made with 
MERLIN, a national facility operated by the University of Manchester at Jodrell Bank Observatory on behalf of PPARC, 
and observations made with the Very Large Array (VLA) of the National Radio Astronomy Observatory (NRAO); the NRAO is 
a facility of the National Science Foundation operated under cooperative agreement by Associated Universities, Inc.

\bibliographystyle{mn2e}

\label{lastpage}
\end{document}